\documentclass[conference]{IEEEtran}

\usepackage{cite}
\usepackage{url}
\usepackage{amsmath,amssymb,amsfonts}
\usepackage{algorithmic}
\usepackage{graphicx}
\usepackage{lipsum}
\usepackage{hhline}
\usepackage{tabularx}
\usepackage{xcolor}
\usepackage{multirow}
\usepackage{subcaption}
\usepackage{dblfloatfix}    
\usepackage[font=small]{caption}

\graphicspath{{graphics/}}

\makeatletter
\def\url@leostyle{%
  \@ifundefined{selectfont}{\def\UrlFont{\sf}}{\def\UrlFont{\small\ttfamily}}}
\makeatother
\DeclareMathOperator*{\argmin}{arg\,min} 

\def\BibTeX{{\rm B\kern-.05em{\sc i\kern-.025em b}\kern-.08em
    T\kern-.1667em\lower.7ex\hbox{E}\kern-.125emX}}
\begin{document}

\title{Towards Scalable Uncertainty Aware DNN-based Wireless Localisation\\
\thanks{}
}

\author{\IEEEauthorblockN{1\textsuperscript{st} Given Name Surname}
\IEEEauthorblockA{\textit{dept. name of organization (of Aff.)} \\
\textit{name of organization (of Aff.)}\\
City, Country \\
email address or ORCID}
\and
\IEEEauthorblockN{2\textsuperscript{nd} Given Name Surname}
\IEEEauthorblockA{\textit{dept. name of organization (of Aff.)} \\
\textit{name of organization (of Aff.)}\\
City, Country \\
email address or ORCID}
\and
\IEEEauthorblockN{3\textsuperscript{rd} Given Name Surname}
\IEEEauthorblockA{\textit{dept. name of organization (of Aff.)} \\
\textit{name of organization (of Aff.)}\\
City, Country \\
email address or ORCID}
}

\author{
    \IEEEauthorblockN{Artan Salihu$^\dagger$$^\ddagger$, Stefan Schwarz$^\dagger$$^\ddagger$ and Markus Rupp$^\dagger$}
    \IEEEauthorblockA{$^\dagger$ Institute of Telecommunications, Technische Universit{\"a}t (TU) Wien\\
                         $^\ddagger$ Christian Doppler Laboratory for Dependable Wireless Connectivity for the Society in Motion \\
                  Email: \{artan.salihu,stefan.schwarz,markus.rupp\}@tuwien.ac.at\\
              }
}

\maketitle
\begin{abstract}
Existing deep neural network (DNN) based wireless localization approaches typically do not capture uncertainty inherent in their estimates. In this work, we propose and evaluate variational and scalable DNN approaches to measure the uncertainty as a result of changing propagation conditions and the finite number of training samples. Furthermore, we show that data uncertainty is sufficient to capture the uncertainty due to non-line-of-sight (NLOS) and, model uncertainty improves the overall reliability. To assess the robustness due to channel conditions and out-of-set regions, we evaluate the methods on challenging massive multiple-input multiple-output (MIMO) scenarios.
\end{abstract}

\section{Introduction}

The ever-increasing demand for location-enabled applications has sharpened the urge for enhanced accuracy and dependability of wireless localization methods in both indoor and outdoor environments. There exist different strategies that exploit wireless signal information to estimate the unknown position of the transmitter, such as received signal strength (RSS), angle-of-arrival (AOA), and time of arrival (TOA) \cite{wen2019survey}. 

More recently, the deployment of massive multiple-input multiple-output (MIMO) technology \cite{marzetta2010noncooperative} in the fifth generation (5G) networks, has encouraged active research in machine learning (ML) for wireless positioning. Due to a high density of antenna elements at the base station (BS), a considerable amount of channel state information (CSI) can be collected. Estimated high-dimensional CSI at a large antenna BS provides fine-grained user equipment (UE) prints; consequently, it reveals spatiotemporal information about the transmitter itself, as well as its surroundings. This information can be harnessed by ML to train a model with CSI samples of known locations. The CSI of the unknown transmitter is then utilized by the model, to infer its position estimate.

Numerous approaches that make use of CSI with machine learning and, in particular deep learning, have recently been proposed \cite{hsieh2019deep,sun2019fingerprint,gante2020deep,salihu2020low}. These algorithms can learn powerful representations that can map high-dimensional CSI into location information. However, despite the evident improvements in localization accuracy, these estimates are taken blindly while failing to give any useful estimates of their predictive uncertainty. Overconfident incorrect predictions in safety-critical applications can have tragic consequences; hence the ability to properly capture and reason about the uncertainty of estimated positions is fundamental to integrate deep neural network (DNN) based methods in wireless localization systems.

Being able to capture uncertainty in location estimates due to changing propagation conditions or insufficient CSI training samples is critical not only for assessing how much we can trust those estimates but also for facilitating active learning and improving the availability of DNN based methods. These are highly desired features for localization approaches applied to real-world and safety-related tasks in railroad transportation, vehicular communications, and assets tracking, to name a few.

In this work, we address these challenges of wireless localization by additionally providing confidence estimates in contrast to only position estimates, which account for both data as well as model uncertainty. However, we aim to learn not only high-accuracy location information but also highlight difficult situations where the model cannot reliably estimate the location of the unknown transmitter. Recently, \cite{wang2021deep} utilizes deep convolutional Gaussian Processes (DCGP) to allow for uncertainty estimation in localization for millimiter Wave communications while improving accuracy. DCGP uses no neural network component. Thus, to the best of our knowledge, this is the first time that DNN uncertainty estimation has been addressed in wireless localization.

We structure the remaining of the paper as follows. In Section \ref{systemModel}, we describe the system model considered for this work. Then, in Section \ref{uePosUncertainty}, we provide details on the proposed approaches for location and uncertainty estimation. In Section \ref{experiments}, we evaluate the localization accuracy and demonstrate the quality of uncertainty estimation for both indoor and outdoor environments. Finally, in Section \ref{conclusion}, we draw our conclusions.

\section{System Model}\label{systemModel}
We consider that the base station receives uplink signal information from $\{\mathbf{x}_{r}\in\mathbb{R}^{d}\}_{r=1}^R$ different locations, corresponding to $R$ single-antenna transmitters, where we consider $d = 2$.
In addition, we assume that users are either in LOS or NLOS with the base station. For the BS, we assume $M = M_yM_z$ antenna elements with $M_y$ and $M_z$ corresponding to the number of antennas in horizontal $(y)$ and vertical $(z)$ directions. The signal from each UE that is received at $M$ antenna elements contains $N_\mathrm{SC}$ subcarriers. Furthermore, due to possible scatterers or reflectors in the reference scenarios, we assume the signal arrives over multiple paths, $L$. Thus, the position-related parameters captured in a scenario-specific multi-path channel for the subcarrier $n$ are given as \vspace{-0.2cm}
\begin{equation}
	\widehat{\mathbf{h}}[n] =\sum_{\ell=1}^{L} \sqrt{\frac{\rho_{\ell}}{N_\mathrm{SC}}} e^{j\frac{2 \pi n}{{N_\mathrm{SC}}} \tau_{\ell} B} \mathbf{a}\left(\varphi_{a z,\ell}, \varphi_{e l,\ell}\right),\vspace{-0.05cm}
\end{equation}
\vspace{-0.05cm}with $\rho_{\ell}$, $B$ and $\tau_{\ell}$ denoting the channel gain, bandwidth and the propagation delay, respectively. The BS steering vector introduces the azimuth and elevation angles of arrivals, $\varphi_{az}$, $\varphi_{el}$, for each path $\ell = 1, \dots, L$,  and is defined as \vspace{-0.2cm}
\begin{equation}
	\mathbf{a}\left(\varphi_{\mathrm{az}}, \varphi_{\mathrm{el}}\right)=\mathbf{a}_{z}\left(\varphi_{\mathrm{el}}\right) \otimes \mathbf{a}_{y}\left(\varphi_{\mathrm{az}}, \varphi_{\mathrm{el}}\right) \vspace{-0.2cm}.
\end{equation}
For a $d=\lambda_{c}/2$ equidistant antenna elements geometry, the array steering vectors $\mathbf{a}_{y}(\cdot), \mathbf{a}_{z}(\cdot)$ in $y$ and $z$ directions are further expressed as \vspace{-0.2cm}
\begin{equation}
	\begin{aligned}
		\mathbf{a}_{y}\left(\varphi_{\mathrm{az}}, \varphi_{\mathrm{el}}\right)=\left[1, e^{j \frac{2\pi}{\lambda_c} d \sin \left(\varphi_{\mathrm{el}}\right) \sin \left(\varphi_{\mathrm{az}}\right)}, \ldots\right.\\ 
		\left.\ldots, e^{j \frac{2\pi}{\lambda_c} d\left(M_{y}-1\right) \sin \left(\varphi_{\mathrm{el}}\right) \sin \left(\varphi_{\mathrm{az}}\right)}\right]^{T},
	\end{aligned}
\end{equation}
\vspace{-0.1cm}
\begin{equation}
	\mathbf{a}_{z}\left(\varphi_{\mathrm{el}}\right)=\left[1, e^{j \frac{2\pi}{\lambda_c} d \cos \left(\varphi_{\mathrm{el}}\right)}, \ldots, e^{j \frac{2\pi}{\lambda_c} d\left(M_{z}-1\right) \cos \left(\varphi_{\mathrm{el}}\right)}\right]^{T}.\vspace{-0.0cm}
\end{equation}
To input the data into the DNN, we handle the complex-valued CSI as two independent real numbers, i.e., $\Re\left\{\widehat{\mathbf{h}}[n]\right\}, \Im\left\{\widehat{\mathbf{h}}[n]\right\}$, representing the real and imaginary components of $\widehat{\mathbf{h}}[n]$. For this paper, we utilize only a single subcarrier for localization and set $n = 1$. Thus, our channel vector is $\mathbf{{h}}\in \mathbb{R}^{2M}$.\vspace{-0.00cm}
\section{Localization with uncertainty estimation}\label{uePosUncertainty}
We use a deep feedforward neural network for position estimation. The network architecture is similar to the \emph{base network} we used in \cite{salihu2020low}. This model consists of a relatively simple and low-complexity feedforward neural network architecture. We define the localization problem as a regression task and the DNN as a function $f_\theta:\mathbb{R}^{2M} \mapsto \mathbb{R}^{d}$ parameterized by $\theta$ where, given the input channel state vector $\mathbf{h}$, we aim to directly map into position-related information, $\mathbf{x}$. Given a training dataset of $N$ i.i.d. sample pairs, $\mathcal{D} = \left\{H, X \right\} =\{\mathbf{h}_n, \mathbf{x}_n\}_{n=1}^N $, the set of optimal parameter values $\theta$ is learned by minimizing a given loss function, $\mathcal{L}\left(\cdot \right) $, \vspace{-0.2cm}
\begin{equation}
	\argmin _{\theta} J(\theta) ; \qquad J(\theta)=\frac{1}{N} \sum_{n=1}^{N} \mathcal{L}(\mathbf{x}_n, \mathbf{h}_n, \theta) \vspace{-0.1cm}
\end{equation}
Usually, the training is performed to minimize the sum of squared errors, $\mathcal{L}(\mathbf{x}_n, \mathbf{h}_n, \theta)=\frac{1}{2}\left\|\mathbf{x}_{n}-f_{\theta}({\mathbf{h}}_{n})\right\|^{2}$. However, such a DNN-based approach leads to a fully deterministic network which outputs only point estimates of the network $\mathbf{x}^{\star}=f_{\theta}\left(\mathbf{h}^{\star}\right)$. This can be interpreted as outputting only the mean of a probability distribution while disregarding other moments. In this paper, however, we aim to provide a probabilistic method that provides not only accurate position estimates of the radio transmitter but also reliable uncertainty associated with the output estimates. To do so, we model the DNN to explicitly learn the underlying uncertainty too. Furthermore, we acquire data and model uncertainty separately. Finally, we also combine both uncertainties to acquire the total uncertainty into one end-to-end model. Next, we describe data uncertainty and in Sec. \ref{section:model_uncertainty} model uncertainty.
\subsection{Data uncertainty} \label{dataUncert}
Since data uncertainty is a property of the data itself, we train the network to directly output the parameters of a probability distribution. To do so, we use a Gaussian mixture model (GMM). In this case, the model yields mixtures of normal distributions, conditioned on the input CSI $\mathbf{h}_n$:\vspace{-0.0cm}
	\begin{equation}
		\begin{array}{c}
		p(\mathbf{x}_n|\mathbf{h}_n; \theta) = \sum_{k=1}^{K} \omega_{\theta,k} \mathcal{N}\left(\mathbf{x}_n ;\boldsymbol{\mu}_{\theta,k}(\mathbf{h}_n), \boldsymbol{\sigma}_{\theta,k}^2(\mathbf{h}_n)\right),
		\end{array}
	\end{equation}
where $K$ is the total number of mixtures and, $\omega_{\theta,k}$, $\boldsymbol{\mu}_{\theta,k}$, and $\boldsymbol{\sigma}_{\theta,k}^2$ being the mixture weight, means, and variances of the $k-$th Gaussian mixture, respectively. We treat $x-$ and $y-$coordinates of $\mathbf{x}$ as independent and restrict to a diagonal covariance matrix. Still, arbitrary distributions can be approximated by using the contribution from multiple mixtures \cite{Bishop94mixturedensity,makansi2019overcoming}. Then, we take a maximum likelihood perspective (MLE) and aim to learn a model that infers the parameters $\boldsymbol{\mu}$ and $\boldsymbol{\sigma^2}$ that maximize the likelihood of observing the desired location, $\mathbf{x}$. This is achieved by minimizing the negative log-likelihood (NLL) as \vspace{-0.0cm}
\begin{equation}
\sum_{n=1}^{N} \underbrace{-\log \sum_{k=1}^{K} \omega_{\theta,k} \mathcal{N}\left(\mathbf{x}_n ;\boldsymbol{\mu}_{\theta,k}(\mathbf{h}_n), \boldsymbol{\sigma}_{\theta,k}^2(\mathbf{h}_n)\right)}_{=: \mathcal{L}(\mathbf{x}_n, \mathbf{h}_n, \theta)}. \label{eqn:NLL}\vspace{-0.0cm}
\end{equation}
The parameters $\{\omega_k, \boldsymbol{\mu}_k, \boldsymbol{\sigma}^{2}_k\}_{k=1}^K$ are the outputs of the network and depend on the input CSI, $\mathbf{h}_n$. These parameters must satisfy certain constraints, which have to be incorporated accordingly in the DNN \cite{Bishop94mixturedensity}. Therefore, in the case of GMM, the last layer of the network outputs the weights, means, and variances as follows. To satisfy $\sum_{k=1}^{K} \omega_k = 1$ and output the probability values corresponding to the weights of the mixture in the range of $0 \leq \omega_k \leq 1$, the output for this part is modelled with $\textit{softmax}$ activation as \vspace{-0.1cm}
\begin{equation}
	\omega_{k}=\frac{\exp {\left(z_{k}^\omega\right)}}{\sum_{k'=1}^{K} \exp \left(z_{k'}^\omega\right)} , \label{eq:weights}\\ 
\end{equation}
where $z_k^\omega$ corresponds to the input of the activation function of the neuron in the output layer for this part.
Likewise, a \textit{softplus} activation function is adopted to satisfy the variance constraint, i.e., $\boldsymbol{\sigma}_{k}^2 \geq 0 , \vspace{-0.1cm}$
\begin{equation}
	\boldsymbol{\sigma}_{k}^2=\log \left(1+\exp \left(\mathbf{z}_{k}^{{\boldsymbol{\sigma}}^2}\right)\right) , \label{eq:softplus} \vspace{-0.1cm}
\end{equation}
where $\mathbf{z}_{k}^{\boldsymbol{\sigma}^2}$ denote the inputs of activation function of units for the part of variance.
For the means, we simply model it using an identity function, i.e., $\mu_k = z_{k}^{\mu}$. Similarly, $\mathbf{z}_{k}^{\boldsymbol{\mu}}$ are the inputs of activation function for each neuron in the output layer for the means. Motivated from the models based on the mixture of experts (MoE), where the $k-$th model is considered an expert for certain input space \cite{ masoudnia2014mixture}, we choose the final estimate as the mean $\widetilde{\boldsymbol{\mu}}_\theta$, and variance $\widetilde{\boldsymbol{\sigma}}^2_\theta$, corresponding to the highest weight mixture, $\max_{k\in K} {\omega_k}$. Here, $\widetilde{\boldsymbol{\sigma}}_{\theta}^2$ corresponds to data uncertainty, $\boldsymbol{\sigma}_{data}^2 = \widetilde{\boldsymbol{\sigma}}_{\theta}^2$. \vspace{-0.0cm}
\subsection{Model uncertainty} \vspace{-0.0cm} \label{section:model_uncertainty}
While modeling the parameters of a distribution function can capture the data uncertainty, this does not allow us to gauge model (epistemic) uncertainty, i.e, the uncertainty over the parameters $\theta$. In order to output the confidence of the model, next we discuss two different approaches.

First, we consider a Bayesian perspective similar to \cite{neal2012bayesian,gal2016dropout, loquercio2020general} to propagate the model uncertainty to the output of the network by placing a distribution over the parameters of the network. In this case, the goal is to utilize the posterior distribution $p(\theta|\mathcal{D})$. We approximate the intractable distribution with Monte Carlo (MC) based methods \cite{srivastava2014dropout,gal2016dropout}. We know from \cite{gal2016dropout} that applying dropout during the test time is equivalent to performing variational inference with a Bernoulli distribution. This approximation is given as
\vspace{-0.0cm}
\begin{equation}
	p(\theta|\mathcal{D}) \approx q(\theta;\Phi) = \mathrm{Bern}(\theta;\Phi) \vspace{-0.0cm} 
\end{equation}
where $\Phi$ is the dropout rate on the network weights at each layer. Thus, we perform $S$ stochastic forward passes with dropout at test time on the same input. The mean, as well as the total variance, are evaluated as:\vspace{-0.0cm}
$$
	\widehat{\boldsymbol{\mu}}^{(\mathrm{MCD})}=\frac{1}{S} \sum_{s=1}^{S} \widetilde{\boldsymbol{\mu}}_{\theta^{(s)}}\left(\mathbf{h}\right),\vspace{-0.0cm}
$$
\begin{equation}\label{eq:mean_and_var_MCD_dropout}
	\widehat{\boldsymbol{\sigma}}_{t}^{2 (\mathrm{MCD})} = \underbrace{\frac{1}{S} \sum_{s=1}^{S}\widetilde{\boldsymbol{\sigma}}_{\theta^{(s)}}^2\left(\mathbf{h}\right)}_{\widehat{\boldsymbol{\sigma}}_{data}^2}+\underbrace{\frac{1}{S}\sum_{s=1}^{S}\left(\widetilde{\boldsymbol{\mu}}_{\theta^{(s)}}\left(\mathbf{h}\right)-\widehat{\boldsymbol{\mu}}^{(\mathrm{MCD})}\right)^2}_{\widehat{\boldsymbol{\sigma}}_{model}^2}.
\end{equation}
In (\ref{eq:mean_and_var_MCD_dropout}), $\widehat{\boldsymbol{\mu}}^{(\mathrm{MCD})}$ refers to the mean location estimate from MC-Dropout and $\widehat{\boldsymbol{\sigma}}_{t}^{2 (\mathrm{MCD})}$ refers to the associated total variance. 

For this Monte Carlo based method, the inference computation time scales linearly with the number of collected weights, $S$. Therefore, we also evaluate another effective alternative to estimate the model uncertainty by sampling from an ensemble of $S$ different neural networks \cite{lakshminarayanan2017simple} trained with $S$ randomly initialized sets of weights of the same network architecture. We refer to this as a deep ensemble network (DEN). Similar to the dropout based approach, we obtain the empirical mean $\widehat{\boldsymbol{\mu}}^{(\mathrm{DEN})}$ and total variance $\widehat{\boldsymbol{\sigma}}_{t}^{2 (\mathrm{DEN})}$ of the distribution of location estimates. While for training we require $S$ different independent trained models and sets of parameters to be stored, we only use a single forward pass during inference.
The methods allow for considering data, model, or jointly both types of uncertainties. The highest weight mixture does not vary with $S$ in this work, and others have negligible weights.

\subsection{Performance Metrics}\label{Performance_Metrics}
We measure the location estimation performance in terms of the root mean squared error (RMSE) defined as
\begin{equation}
\mathrm{RMSE}=\sqrt{\frac{\sum_{n=1}^{N_{\mathrm{test}}}\left\|\mathbf{x}_{n}^{\star}-\widehat{\boldsymbol{\mu}}_{n}^{(\cdot)}\right\|^{2}}{N_{\mathrm{test}}} },
\end{equation}
where $\mathbf{x}_{n}^{\star}$ is the actual position of the test location $n$ and, $\widehat{\boldsymbol{\mu}}_{n}^{(\cdot)}$ is the estimated location given the evaluated method, i.e., dropout- or ensemble-based one. For example,  $\widehat{\boldsymbol{\mu}}^{(\cdot)} = \widehat{\boldsymbol{\mu}}^{(\mathrm {MCD})}$ for dropout and $\widehat{\boldsymbol{\mu}}^{(\cdot)} = \widehat{\boldsymbol{\mu}}^{(\mathrm {DEN})}$ for ensemble.

To evaluate the quality of uncertainty estimation, we assess the ordering defined by uncertainty estimates (confidence) \cite{ilg2018uncertainty} compared to the ground-truth error (oracle). Intuitively, removing locations with high uncertainty should lead to lower $\mathrm{RMSE}$. Therefore, we evaluate their difference, i.e., the error between the ordering of locations defined by $\mathrm{RMSE}$ (\textit{oracle}) and the ordering defined by the uncertainty estimates (\textit{confidence}),
\begin{equation}
\alpha_i = \mathrm{RMSE}_\mathrm{orac}(b_i) - \mathrm{RMSE}_\mathrm{conf}(b_i),
\end{equation}
where $b_i$ represents the fraction of removed locations. Furthermore, to compare the two methods for different numbers of ensembles and MC-dropout forward passes with a single value, we evaluate the area under the confidence-oracle error curve, denoted as $\mathrm{AUCO}$.
The smaller $\mathrm{AUCO}$ value, the better acquired uncertainty explains the variations in locations with respect to $\mathrm{RMSE}$.

\section{Experiments and Results}\label{experiments}
In this section we describe the parameter details for investigated scenarios, training details, and the results for performance investigation for both indoor and outdoor environments.\vspace{-0.0cm}
\subsection{Simulation parameters}\label{sec:simulation_parameters}
We evaluate the proposed approaches on two ray-tracing based outdoor and indoor scenarios \cite{alkhateeb2019deepmimo}:
\begin{itemize}
	\item[\textit{1)}] \textit{The indoor scenario} considered is denoted as $\textrm{I3\_2p4}$. This is a scenario with mixed user locations in LOS and others in NLOS.
	\item[\textit{2)}] \textit{The outdoor scenario} of our interest is $\textrm{O1\_3p5B}$. Similarly, this scenario has LOS as well as NLOS user locations blocked by a metal screen which is placed in front of the BS. In addition, two reflecting surfaces for the NLOS users to the BS are also present.
\end{itemize}                      
We consider $L=5$ paths and a uniform planar array at the BS with $M = M_y \times M_z = 16 \times 8$. The region considered for $\textrm{O1\_3p5B}$ is R$800-$R$1200$, i.e., the rows in a grid layout. Each row has $R'=181$ user locations and all users in this region are served by BS$-3$. Table \ref{table: scenario parameters} summarizes the simulation parameters.
\begin{table}[!h]
	\caption{Parameters for the investigated scenarios.}
	\begin{tabularx}{\columnwidth}{X||X|X}
		\textbf{ }                     & \textbf{Scenario I3\_2p4}  & \textbf{Scenario O1\_3p5B}   \\ \hhline{=||=|=}
		\multicolumn{1}{c||}{Frequency, $\mathit{f}_{c}$}     & \multicolumn{1}{c|}{2.4 GHz} & \multicolumn{1}{c}{3.5 GHz}  \\ \hline
		\multicolumn{1}{c||}{Bandwidth, $B$}      & \multicolumn{1}{c|}{20 MHz}  & \multicolumn{1}{c}{20 MHz}   \\ \hline
		\multicolumn{1}{c||}{BS Number}      & \multicolumn{1}{c|}{BS$-$2}    & \multicolumn{1}{c}{BS$-$3}     \\ \hline
		\multicolumn{1}{c||}{Numer of paths, $L$} & \multicolumn{1}{c|}{5}       & \multicolumn{1}{c}{5}        \\ \hline
		\multicolumn{1}{c||}{Subcarriers, $N_\mathrm{SC}$}  & \multicolumn{1}{c|}{1024}    & \multicolumn{1}{c}{1024}     \\ \hline
		\multicolumn{1}{c||}{User locations} & \multicolumn{1}{c|}{R$1-$R$1159$}  & \multicolumn{1}{c}{R$800-$R$1200$} \\ \hline \hline
	\end{tabularx}
	\label{table: scenario parameters} \vspace{-0.2cm}
\end{table}
\begin{figure}[!t]
	\setlength\belowcaptionskip{0pt} 
	\centering
	\label{Network_architecture}{%
		\includegraphics[width=0.90\linewidth]{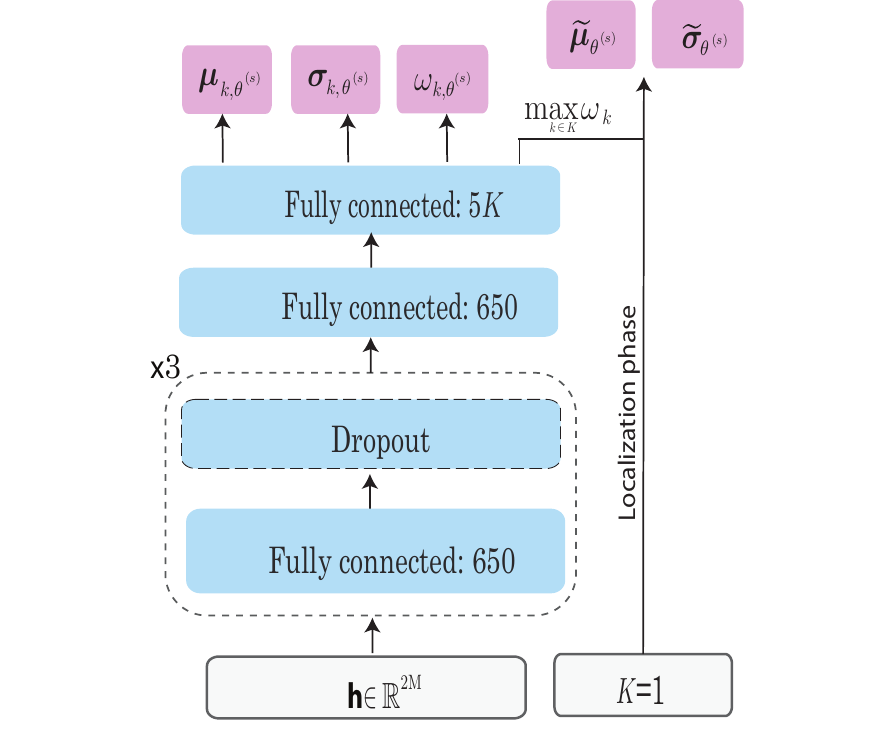}}\vspace{0.2cm} 
	\caption{Model architecture overview. Given the CSI, the network learns a full parametric Gaussian mixture model (GMM) over locations. During localization phase, the optimal location estimate is considered from highest weight mixture with the associated variance.}
	\label{fig:Network_architecture}\vspace{-0.1cm}
\end{figure}

\begin{figure}[!t] 
	\centering
	\subfloat[Indoors $(S=32)$\label{fig:ecdf_indoor}]{%
		\includegraphics[width=0.49\linewidth]{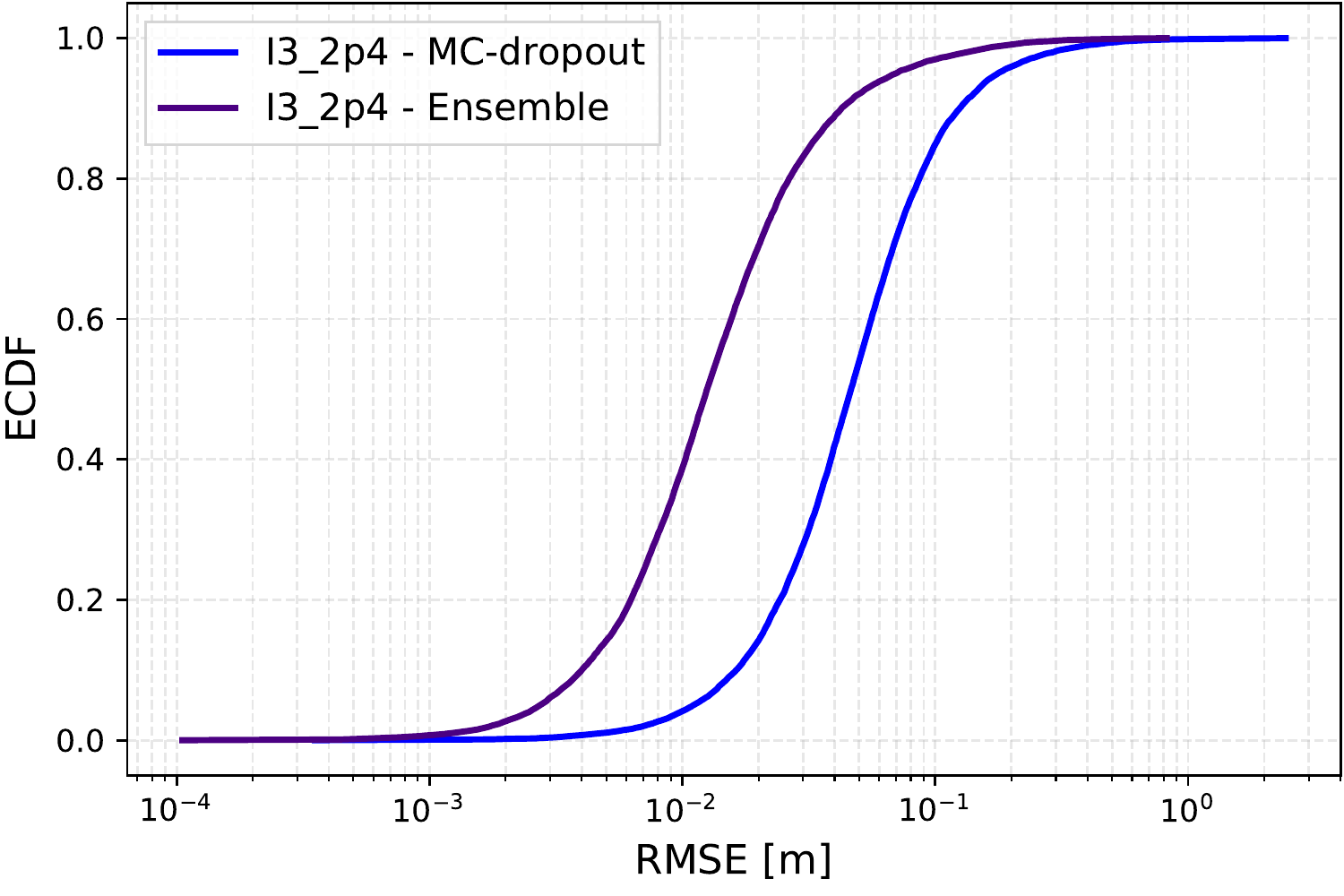}}
	\hfill
	\subfloat[Outdoors $(S=32)$ \label{fig:ecdf_outdoor}]{%
		\includegraphics[width=0.5\linewidth]{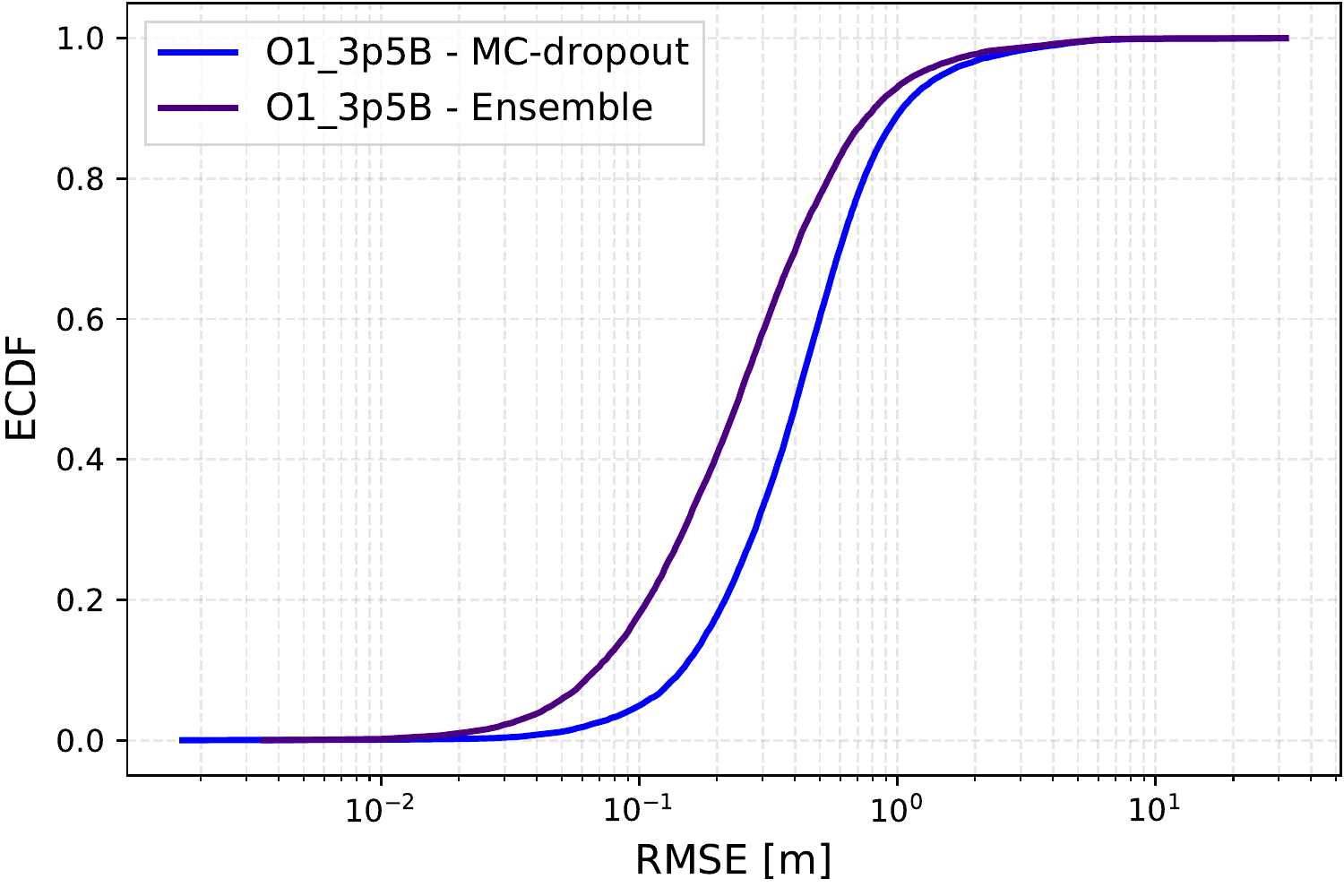}}
	\\
	\subfloat[Indoors\label{fig:rmse_indoor}]{%
		\includegraphics[width=0.5\linewidth]{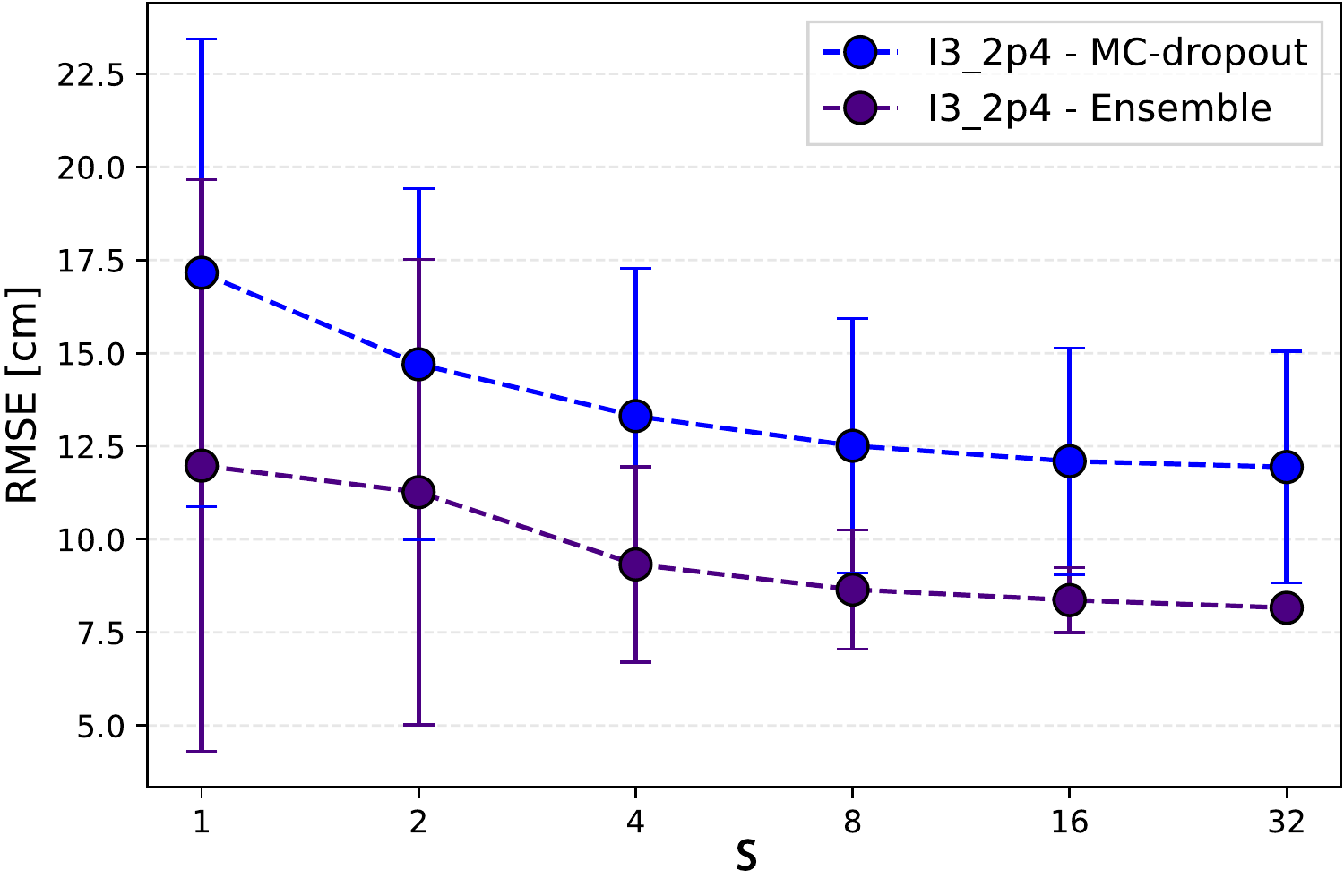}}
	\hfill
	\subfloat[Outdoors\label{fig:rmse_outdoor}]{%
		\includegraphics[width=0.5\linewidth]{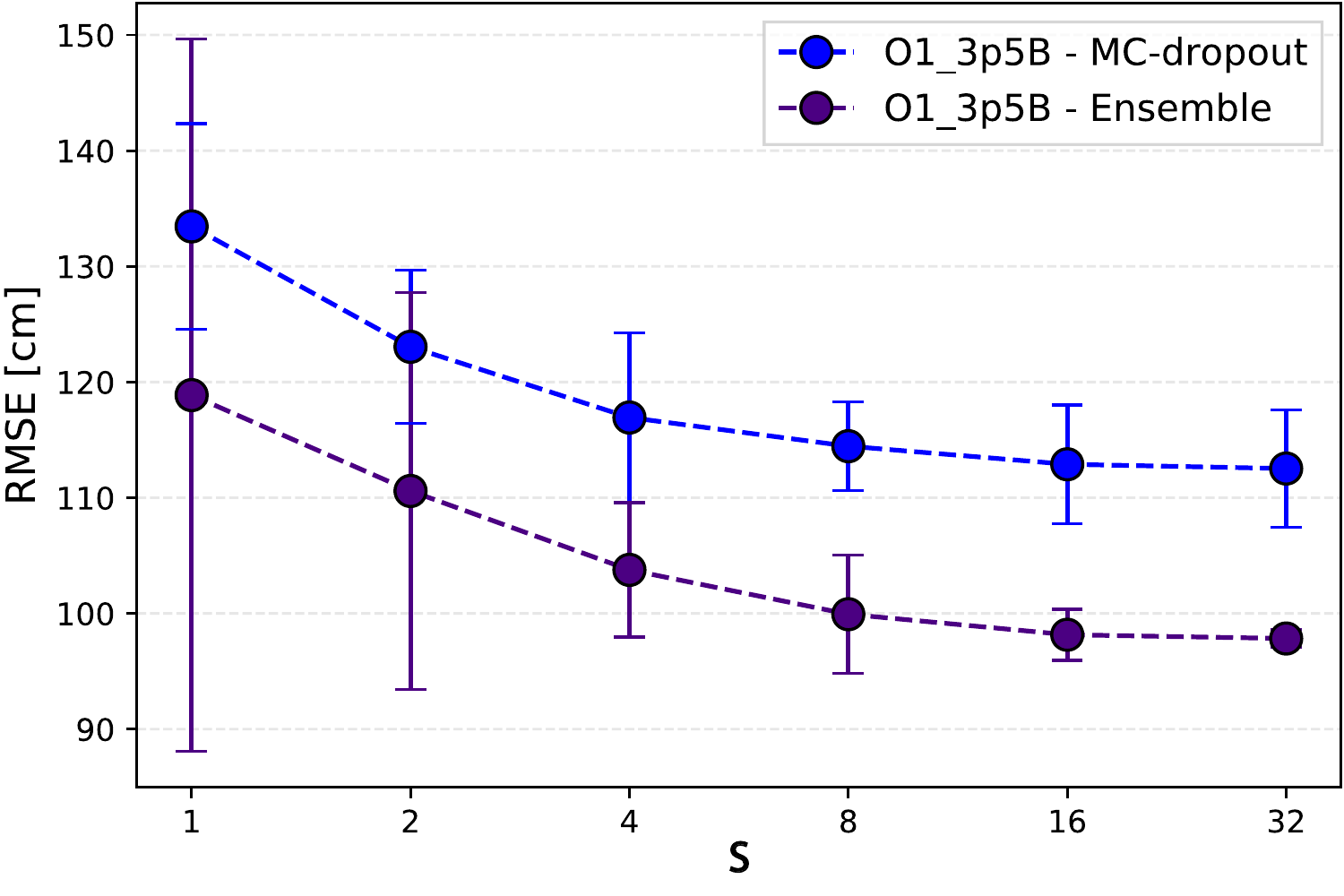}}
	
	\caption{Localization error for indoors (a) and outdoors (b) ($S=32$).  Accuracy improves for $S>1$ for both indoors (c) and outdoors (d) scenarios. $\mathrm{DEN}$ performs better than $\mathrm{MCD}$ in terms of $\mathrm{RMSE}$.}
	\label{fig:position_accuracy_with_cdf}\vspace{-0.7cm}
\end{figure}
\subsection{Training Details}
The network architecture is composed of $V=4$ hidden layers, as illustrated in Fig.~\ref{fig:Network_architecture}. We adopt $\mathrm{ReLU}$ for the layers $v =\{1,2,3,4\}$. We model the output layer of the network as described in Section \ref{dataUncert}, which outputs $\widehat{\boldsymbol{\mu}}_{k}^{(\cdot)}$, $\widehat{\boldsymbol{\sigma}}_{k}^{(\cdot)}$ and $\widehat{\omega}_{k}^{(\cdot)}$. The size of the output layer is $ 5K = 15$ units, i.e., neurons.
For MC-Dropout, we place the dropout layer after $v = \{1, 2, 3\}$ of the network and search over a grid for $\Phi \in \{0.05, 0.1, 0.2\}$. For the presented experiments, a dropout rate of $0.1$ was selected. We train the model for $600$ epochs with Adam \cite{ruder2016overview}, batch size of $512$ at a fixed learning rate of $10^{-3}$, and early stopping if validation loss is not reduced for $80$ consecutive epochs. Weights are initialized from $\mathcal{N}(0,10^{-2})$. For the ensemble approach, we train all individual networks for $300$ epochs without dropout, thus faster converge time. However, to regularize the training process, in addition to early-stopping after $30$ epochs, we also clip the gradients at a value of $1.0$. Other parameters are kept the same as for MC-dropout. The models are trained in Tensorflow \cite{abadi2016tensorflow}.

Finally, to facilitate the training convergence time for these scenarios, we scale the dataset by dividing the inputs with the maximum absolute value in the dataset, $\Delta_{\text {norm }} = \max(\{|{h}_{n,1}|, \ldots, |{h}_{n,2M}|\}_{n=1}^{N})$ \cite{alkhateeb2018deep}. Similarly, the position coordinate values are scaled in the range of $[0,1]$. However, during the testing phase, the estimates are reverted to the original scale to evaluate the performance in terms of $\mathrm{RMSE}$.
\begin{figure}[!t]
	\centering
	\subfloat[Confidence-oracle curve\label{fig:confidence_oracle_curve}]{%
		\includegraphics[width=0.5\linewidth]{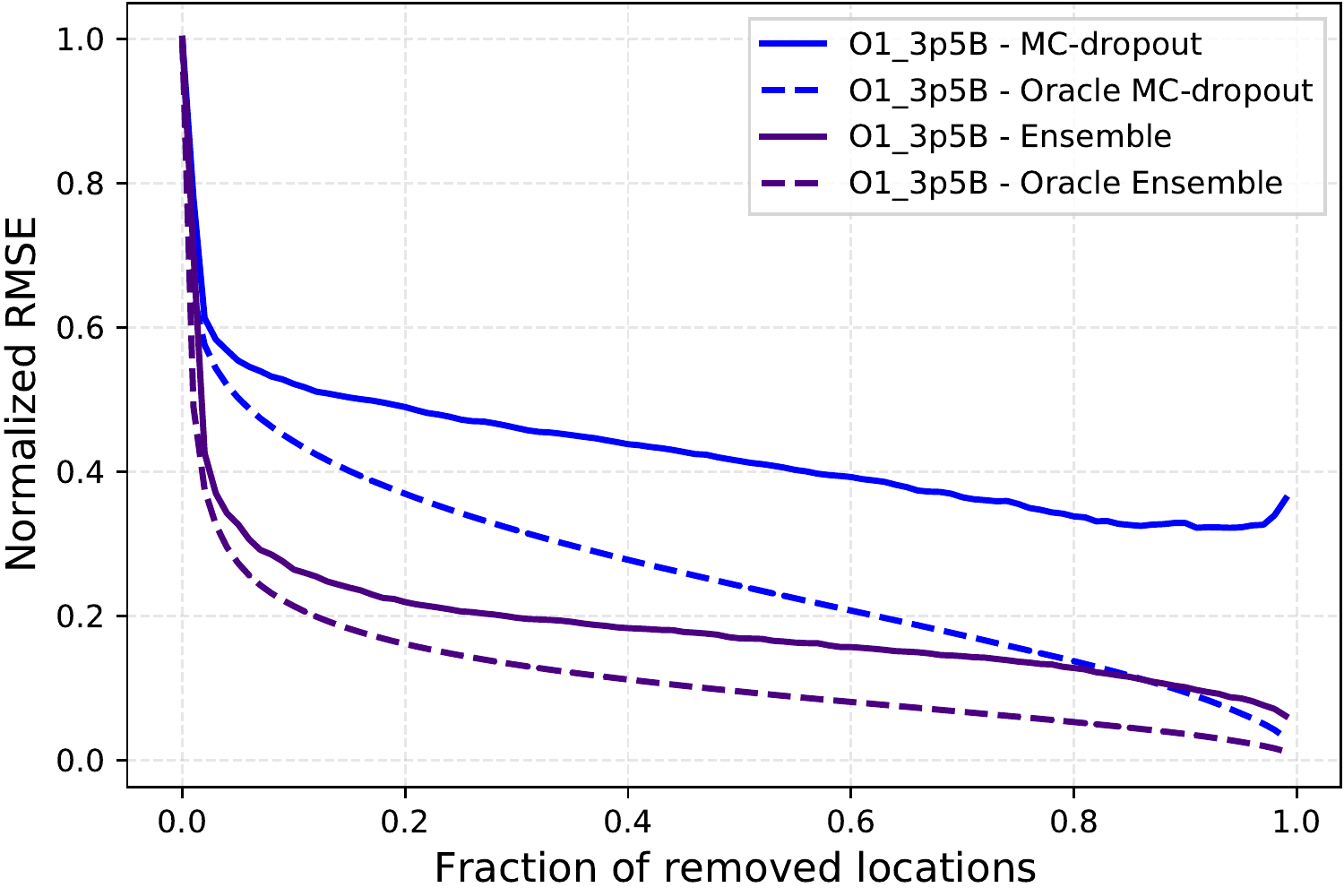}}
	\hfill
	\subfloat[Confidence-oracle error\label{fig:confidence_oracle_error}]{%
		\includegraphics[width=0.5\linewidth]{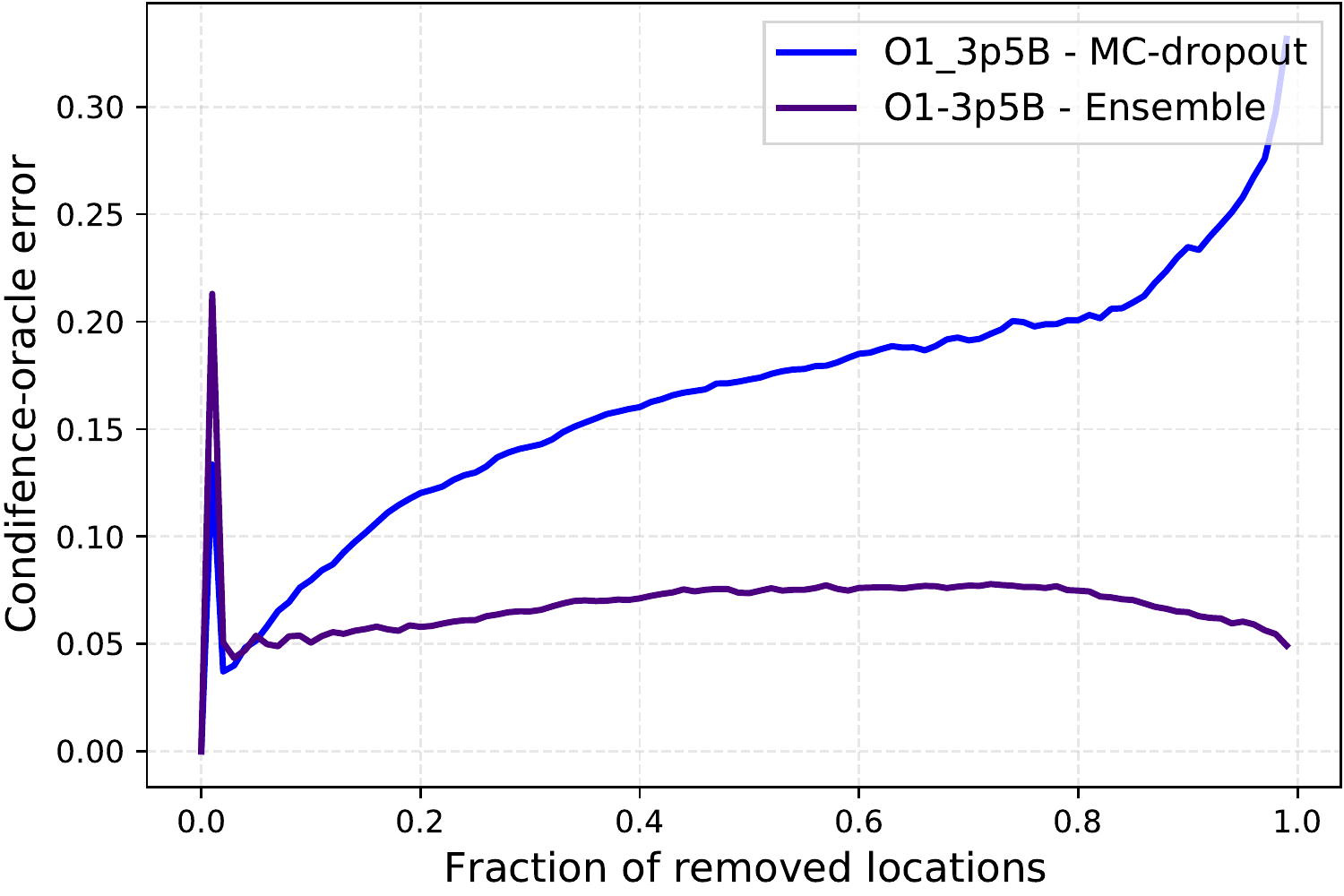}}
	\\
	\subfloat[Indoors \label{fig:aucoIndoor}]{%
		\includegraphics[width=0.5\linewidth]{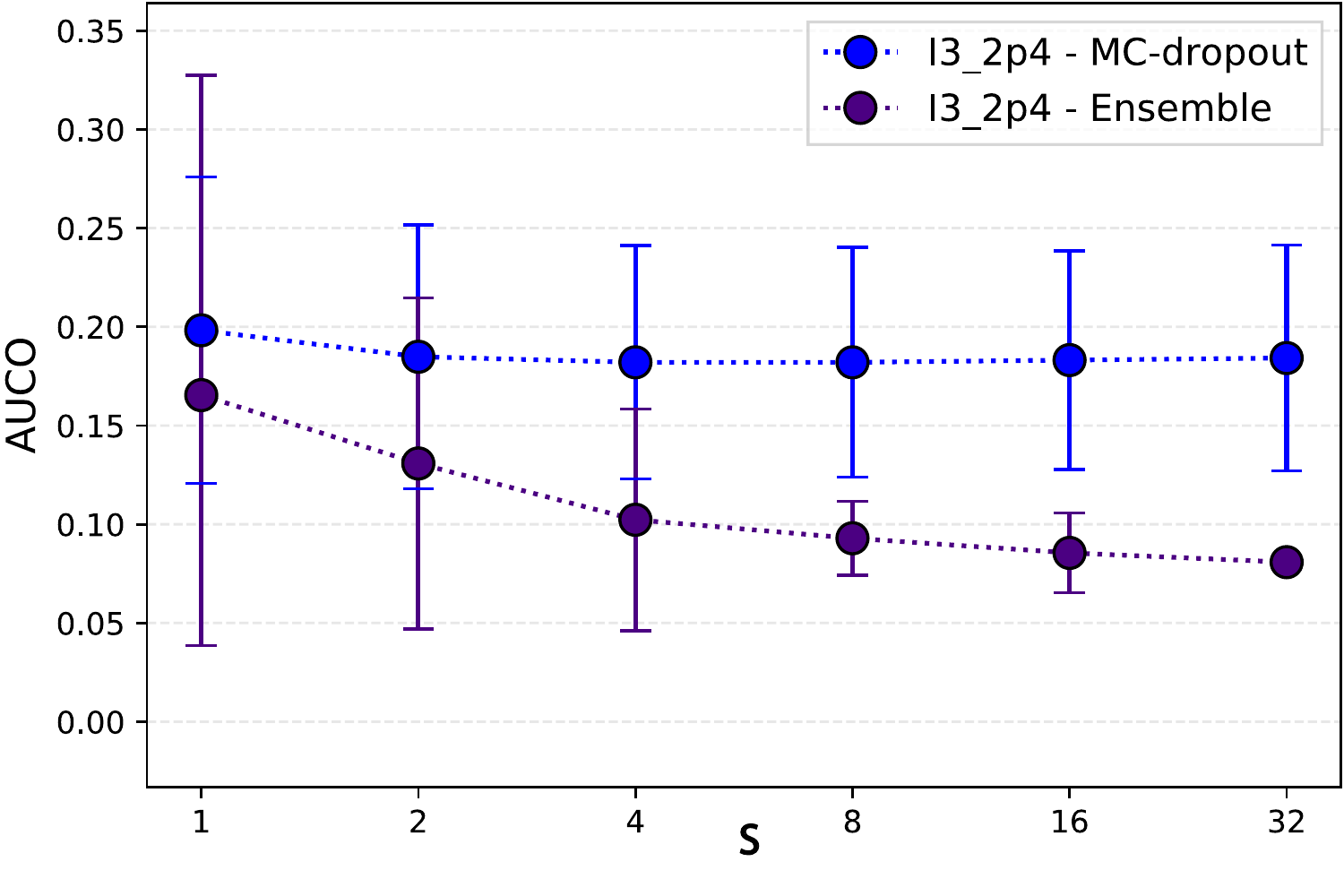}}
	\hfill
	\subfloat[Outdoors \label{fig:aucoOutdoor}]{%
		\includegraphics[width=0.5\linewidth]{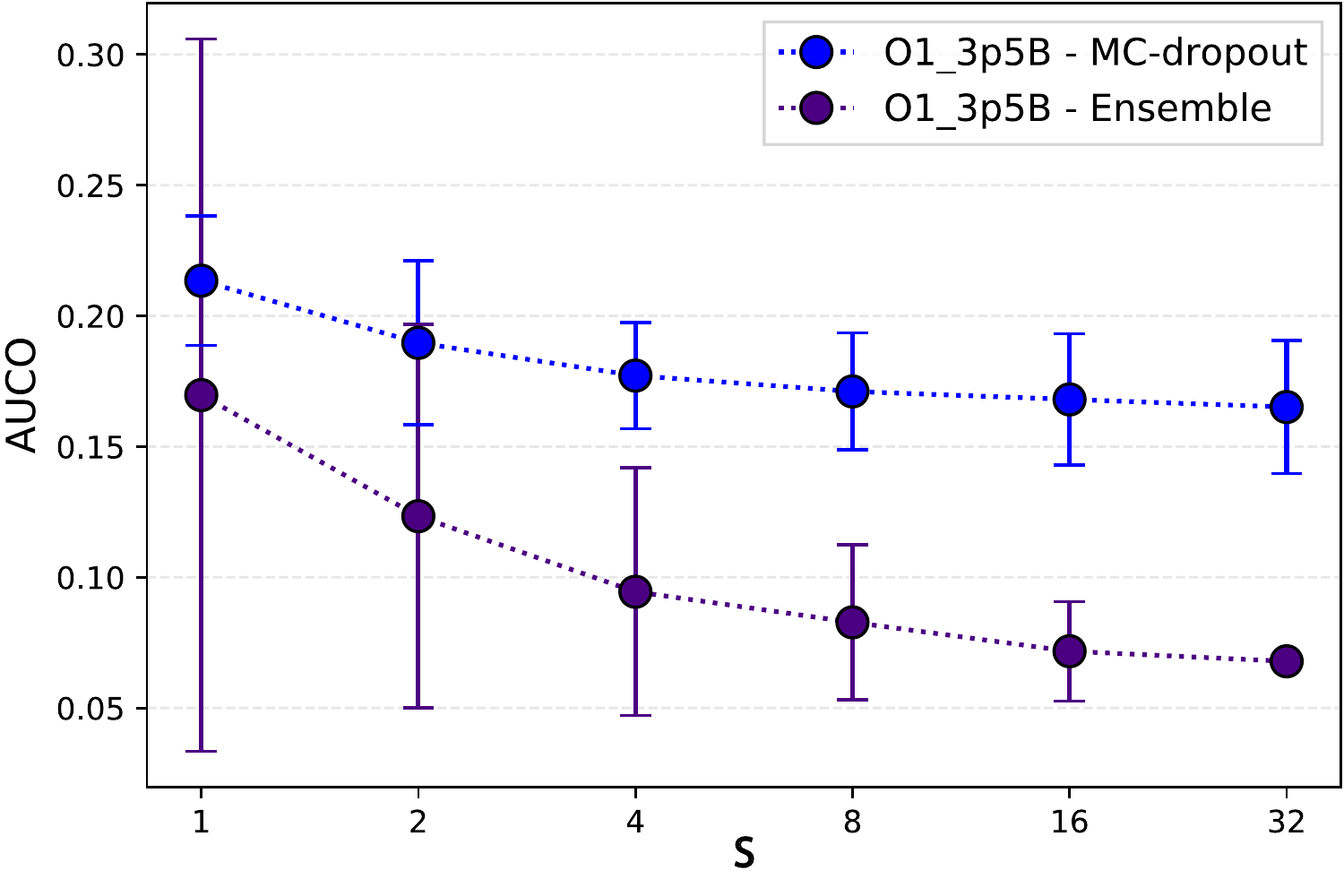}}
	\vspace{0.08cm}
	\caption{Variance based ordering evaluation for uncertainty estimation. Example of confidence and oracle curves (a) and error between the confidence and oracle curve (b) for $S = 32$. $\mathrm{DEN}$ outperforms $\mathrm{MCD}$ in terms of AUCO for both indoors (c) and outdoors (d) scenarios.}
	\label{fig:rankingMetrics} \vspace{-0.3cm}
\end{figure}
\vspace{-0.4cm}\subsection{Localization accuracy}
Localization accuracy in terms of $\mathrm{RMSE}$ for the two approaches and all reference scenarios are depicted in Fig.~\ref{fig:position_accuracy_with_cdf}. We observe in Fig.~\ref{fig:rmse_indoor}), \ref{fig:rmse_outdoor}) that accuracy improves with $S$. Averaging over $S$ different weight configurations has a pronounced positive impact on the overall $\mathrm{RMSE}$. For the sake of comparison, we also provide results for $S=1$. This is equivalent to only estimating $\boldsymbol{\sigma}_{data}^2$. Finally, we can observe that $\mathrm{DEN}$ outperforms $\mathrm{MCD}$.\vspace{-0.0cm}
\subsection{Uncertainty accuracy}
Fig.~\ref{fig:confidence_oracle_curve} and Fig.~\ref{fig:confidence_oracle_error} show that a deep ensemble can better capture the variations in the RMSE. The evaluation in terms of $\mathrm{AUCO}$ is depicted in  Fig.~\ref{fig:aucoIndoor} and  Fig.~\ref{fig:aucoOutdoor} for indoor and outdoor scenario respectively. We can easily notice that the ensemble performs better than MC-dropout and the gap is more evident as $S>2$. We can also observe in Fig.~\ref{fig:confidence_oracle_curve} that removing $20\%$ of locations with the highest error, the overall accuracy improves by $80\%$ for an ensemble-based approach in this outdoor scenario.\vspace{-0.0cm}
\subsection{Impact of data and model uncertainty estimation}
\begin{figure}[!t]
	\centering
	\subfloat[MC-dropout ($\mathrm{MCD}$) \label{heatmapMC}]{%
		\includegraphics[width=0.97\linewidth]{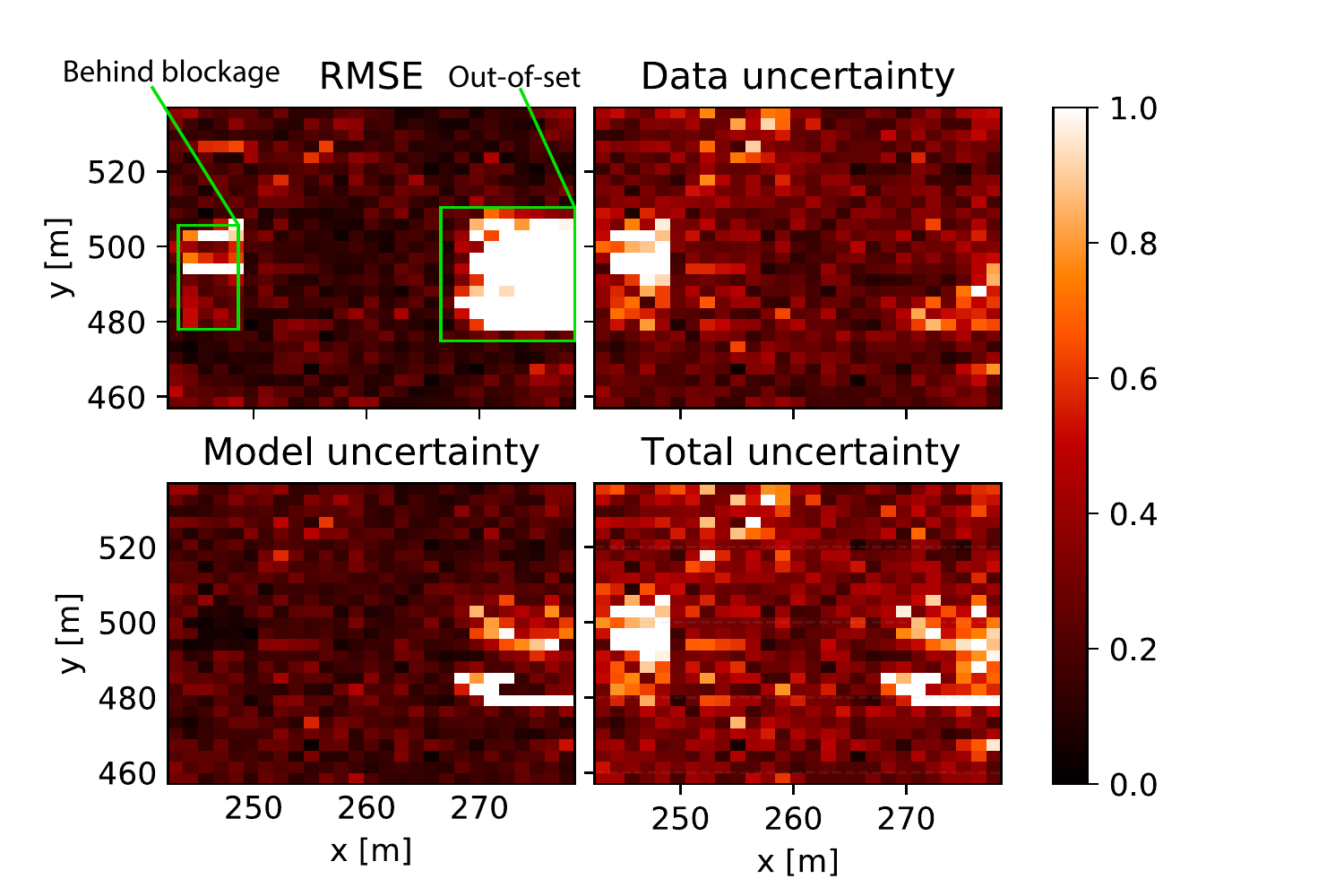}} \\
	\vspace{-0.0cm}
	\subfloat[Ensemble ($\mathrm{DEN}$) \label{heatmapEnsemble}]{%
		\includegraphics[width=0.97\linewidth]{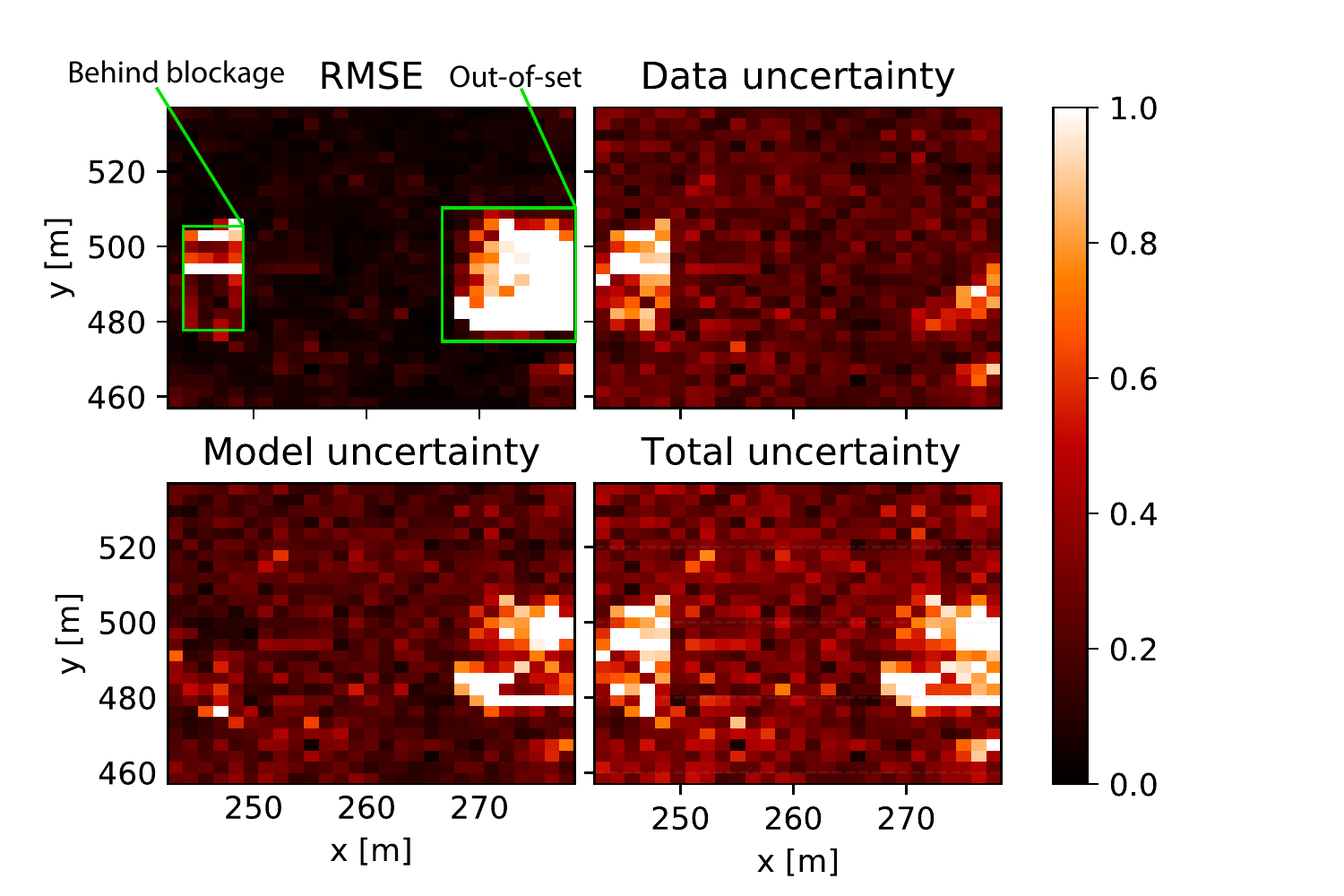}}
	\caption{The impact of data and model uncertainty estimation in NLOS and out-of-set case. Lighter colored regions indicate higher error and uncertainty. $\mathrm{DEN}$ (b) better represents the error in out-of-set region.}
	\label{fig:heatmapsO1_3p5}\vspace{-0.4cm}
\end{figure}

In Fig.~\ref{fig:heatmapsO1_3p5}, we provide qualitative results for uncertainty estimation for the two cases: NLOS and out-of-set region. To do so, we consider the outdoor scenario $\textrm{O1\_3p5}$, described in Sec. \ref{sec:simulation_parameters}, where locations for users behind the blockage have the highest $\mathrm{RMSE}$; consequently, both approaches should provide high uncertainty too. Moreover, our goal is also to understand if model uncertainty can improve our awareness about out-of-set regions. Therefore, we remove all training samples from a region marked in a green rectangle box as out-of-set region in Fig.~\ref{fig:heatmapsO1_3p5}. Likewise, we expect the methods to estimate high uncertainty for users in the out-of-set region during the testing phase. We show that data uncertainty is sufficient to capture the error due to NLOS. However, out-of-set cases are more challenging and acquiring model uncertainty enhances the overall dependability.\vspace{0.2cm}
\vspace{-0.2cm}
\section{Conclusion}\label{conclusion}
\vspace{-0.1cm}
In this work, we addressed and investigated a fundamental issue in current DNN-based wireless localization methods, i.e., uncertainty unawareness due to propagation conditions and insufficient training samples. We proposed and evaluated scalable DNN based approaches that can implicitly learn and output the uncertainty in addition to accurate position estimates for wireless localization. We also evaluated the quality and showed that data uncertainty is sufficient to capture uncertainty due to NLOS. Finally, we showed that model uncertainty improves the reliability when the DNN operates in an out-of-set region, especially for the deep ensemble network.
\vspace{-0.0cm}
\small \section*{Acknowledgment}
\vspace{0.1cm}
This work has been funded by \"OBB Infrastruktur AG. The financial support by the Austrian Federal Ministry for Digital and Economic Affairs and the National Foundation for Research, Technology and Development is gratefully acknowledged.
\vspace{0.4cm}
\bibliographystyle{IEEEtran}
\bibliography{References}

\begin{thebibliography}{10}
\providecommand{\url}[1]{#1}
\csname url@samestyle\endcsname
\providecommand{\newblock}{\relax}
\providecommand{\bibinfo}[2]{#2}
\providecommand{\BIBentrySTDinterwordspacing}{\spaceskip=0pt\relax}
\providecommand{\BIBentryALTinterwordstretchfactor}{4}
\providecommand{\BIBentryALTinterwordspacing}{\spaceskip=\fontdimen2\font plus
\BIBentryALTinterwordstretchfactor\fontdimen3\font minus
  \fontdimen4\font\relax}
\providecommand{\BIBforeignlanguage}[2]{{%
\expandafter\ifx\csname l@#1\endcsname\relax
\typeout{** WARNING: IEEEtran.bst: No hyphenation pattern has been}%
\typeout{** loaded for the language `#1'. Using the pattern for}%
\typeout{** the default language instead.}%
\else
\language=\csname l@#1\endcsname
\fi
#2}}
\providecommand{\BIBdecl}{\relax}
\BIBdecl

\bibitem{wen2019survey}
F.~Wen, H.~Wymeersch, B.~Peng, W.~P. Tay, H.~C. So, and D.~Yang, ``A survey on
  {5G} massive {MIMO} localization,'' \emph{Digital Signal Processing},
  vol.~94, pp. 21--28, 2019.

\bibitem{marzetta2010noncooperative}
T.~L. Marzetta, ``Noncooperative cellular wireless with unlimited numbers of
  base station antennas,'' \emph{IEEE transactions on wireless communications},
  vol.~9, no.~11, pp. 3590--3600, 2010.

\bibitem{hsieh2019deep}
C.-H. Hsieh, J.-Y. Chen, and B.-H. Nien, ``Deep learning-based indoor
  localization using received signal strength and channel state information,''
  \emph{IEEE Access}, vol.~7, pp. 33\,256--33\,267, 2019.

\bibitem{sun2019fingerprint}
X.~Sun, C.~Wu, X.~Gao, and G.~Y. Li, ``Fingerprint-based localization for
  massive {MIMO-OFDM} system with deep convolutional neural networks,''
  \emph{IEEE Transactions on Vehicular Technology}, vol.~68, no.~11, pp.
  10\,846--10\,857, 2019.

\bibitem{gante2020deep}
J.~Gante, G.~Falcao, and L.~Sousa, ``Deep learning architectures for accurate
  millimeter wave positioning in 5{G},'' \emph{Neural Processing Letters},
  vol.~51, no.~1, pp. 487--514, 2020.

\bibitem{salihu2020low}
A.~Salihu, S.~Schwarz, A.~Pikrakis, and M.~Rupp, ``Low-dimensional
  representation learning for wireless {CSI}-based localisation,'' in
  \emph{16th International Conference on Wireless and Mobile Computing,
  Networking and Communications (WiMob)(50308)}.\hskip 1em plus 0.5em minus
  0.4em\relax IEEE, 2020, pp. 1--6.

\bibitem{wang2021deep}
X.~Wang, M.~Patil, C.~Yang, S.~Mao, and P.~A. Patel, ``Deep convolutional
  {Gaussian Processes for Mmwave} outdoor localization,'' in \emph{ICASSP
  2021-2021 IEEE International Conference on Acoustics, Speech and Signal
  Processing (ICASSP)}.\hskip 1em plus 0.5em minus 0.4em\relax IEEE, 2021, pp.
  8323--8327.

\bibitem{Bishop94mixturedensity}
C.~M. Bishop, ``Mixture density networks,'' Tech. Rep., 1994.

\bibitem{makansi2019overcoming}
O.~Makansi, E.~Ilg, O.~Cicek, and T.~Brox, ``Overcoming limitations of mixture
  density networks: A sampling and fitting framework for multimodal future
  prediction,'' in \emph{Proceedings of the IEEE/CVF Conference on Computer
  Vision and Pattern Recognition}, 2019, pp. 7144--7153.

\bibitem{masoudnia2014mixture}
S.~Masoudnia and R.~Ebrahimpour, ``Mixture of experts: a literature survey,''
  \emph{Artificial Intelligence Review}, vol.~42, no.~2, pp. 275--293, 2014.

\bibitem{neal2012bayesian}
R.~M. Neal, \emph{{Bayesian} learning for neural networks}.\hskip 1em plus
  0.5em minus 0.4em\relax Springer Science \& Business Media, 2012, vol. 118.

\bibitem{gal2016dropout}
Y.~Gal and Z.~Ghahramani, ``Dropout as a {Bayesian} approximation: Representing
  model uncertainty in deep learning,'' in \emph{international conference on
  machine learning}, 2016, pp. 1050--1059.

\bibitem{loquercio2020general}
A.~Loquercio, M.~Segu, and D.~Scaramuzza, ``A general framework for uncertainty
  estimation in deep learning,'' \emph{IEEE Robotics and Automation Letters},
  vol.~5, no.~2, pp. 3153--3160, 2020.

\bibitem{srivastava2014dropout}
N.~Srivastava, G.~Hinton, A.~Krizhevsky, I.~Sutskever, and R.~Salakhutdinov,
  ``Dropout: a simple way to prevent neural networks from overfitting,''
  \emph{The journal of machine learning research}, vol.~15, no.~1, pp.
  1929--1958, 2014.

\bibitem{lakshminarayanan2017simple}
B.~Lakshminarayanan, A.~Pritzel, and C.~Blundell, ``Simple and scalable
  predictive uncertainty estimation using deep ensembles,'' in \emph{Advances
  in neural information processing systems}, 2017, pp. 6402--6413.

\bibitem{ilg2018uncertainty}
E.~Ilg, O.~Cicek, S.~Galesso, A.~Klein, O.~Makansi, F.~Hutter, and T.~Brox,
  ``Uncertainty estimates and multi-hypotheses networks for optical flow,'' in
  \emph{Proceedings of the European Conference on Computer Vision (ECCV)},
  2018, pp. 652--667.

\bibitem{alkhateeb2019deepmimo}
A.~Alkhateeb, ``{DeepMIMO}: A generic deep learning dataset for millimeter wave
  and massive {MIMO} applications,'' \emph{arXiv preprint arXiv:1902.06435},
  2019.

\bibitem{ruder2016overview}
S.~Ruder, ``An overview of gradient descent optimization algorithms,''
  \emph{arXiv preprint arXiv:1609.04747}, 2016.

\bibitem{abadi2016tensorflow}
M.~Abadi, A.~Agarwal, P.~Barham, E.~Brevdo, Z.~Chen, C.~Citro, G.~S. Corrado,
  A.~Davis, J.~Dean, M.~Devin \emph{et~al.}, ``Tensorflow: Large-scale machine
  learning on heterogeneous distributed systems,'' \emph{arXiv preprint
  arXiv:1603.04467}, 2016.

\bibitem{alkhateeb2018deep}
A.~Alkhateeb, S.~Alex, P.~Varkey, Y.~Li, Q.~Qu, and D.~Tujkovic, ``Deep
  learning coordinated beamforming for highly-mobile millimeter wave systems,''
  \emph{IEEE Access}, vol.~6, pp. 37\,328--37\,348, 2018.

\end{thebibliography}

\end{document}